\documentstyle[twoside,12pt]{article}

\catcode`\@=11
\long\def\@makefntext#1{ 
\protect\noindent \hbox to 3.2pt {\hskip-.9pt
$^{{\eightrm\@thefnmark}}$\hfil}#1\hfill} 

\def\thefootnote{\fnsymbol{footnote}}
 \def\@makefnmark{\hbox to 0pt{$^{\@thefnmark}$\hss}}  

\def\ps@myheadings{\let\@mkboth\@gobbletwo
\def\@oddhead{\hbox{} 
\rightmark\hfil\eightrm\thepage}
\def\@oddfoot{}\def\@evenhead{\eightrm\thepage\hfil 
\leftmark\hbox{}}\def\@evenfoot{}
\def\sectionmark##1{}\def\subsectionmark##1{}}

\oddsidemargin=\evensidemargin
\renewcommand{\thefootnote}{\fnsymbol{footnote}}

\newcounter{sectionc}\newcounter{subsectionc}\newcounter{subsubsectionc}
\renewcommand{\section}[1] {\vspace{12pt}\addtocounter{sectionc}{1}
\setcounter{subsectionc}{0}\setcounter{subsubsectionc}{0}\noindent
        {\bf\thesectionc. #1}\par\vspace{5pt}}
\renewcommand{\subsection}[1] {\vspace{12pt}\addtocounter{subsectionc}{1}
        \setcounter{subsubsectionc}{0}\noindent
        {\bf\thesectionc.\thesubsectionc. {\kern1pt \bf\it #1}}\par
        \vspace{5pt}}
\renewcommand{\subsubsection}[1] {\vspace{12pt}\addtocounter{subsubsectionc}{1}
        \noindent{\thesectionc.\thesubsectionc.\thesubsubsectionc.
        {\kern1pt \it #1}}\par\vspace{5pt}}
\newcommand{\nonumsection}[1] {\vspace{12pt}\noindent{\bf #1}
        \par\vspace{5pt}}

\newcommand{\textlineskip}{\baselineskip=14pt}

\def\eightcirc{
\begin{picture}(0,0)
\put(4.4,1.8){\circle{6.5}}
\end{picture}}
\def\eightcopyright{\eightcirc\kern2.7pt\hbox{\eightrm c}}

\def\abstracts#1#2#3{{
        \centering{\begin{minipage}{5in}\baselineskip=12pt\tenrm
        \centerline{ABSTRACT}
        \parindent=0pt #1\par
        \parindent=15pt #2\par
        \parindent=15pt #3
        \end{minipage} }\par}}

\renewenvironment{thebibliography}[1]                   
        {
         \begin{list}{\arabic{enumi}.}                  
        {\usecounter{enumi}\setlength{\parsep}{0pt}
         \setlength{\leftmargin 17pt}{\rightmargin 0pt} 
         \setlength{\itemsep}{0pt} \settowidth          
        {\labelwidth}{#1.}\sloppy}}{\end{list}} 

\newcounter{itemlistc}
\newcounter{romanlistc}
\newcounter{alphlistc}
\newcounter{arabiclistc}

\newcounter{tempfigtabc}                        
\newcommand{\fcaption}[1]{
        {\begin{center}{\tenrm Fig.~\thetempfigtabc. #1}\end{center}
        }}

\def\pmb#1{\setbox0=\hbox{#1}
        \kern-.025em\copy0\kern-\wd0
        \kern.05em\copy0\kern-\wd0
        \kern-.025em\raise.0433em\box0}

\def\fnt#1#2{\footnotetext{\kern-.3em
        {$^{\mbox{\scriptsize #1}}$}{#2}}}

 1
 1
 1

\font\eightrm=cmr8

\def\qed{\hbox{${\vcenter{\vbox{                          
   \hrule height 0.4pt\hbox{\vrule width 0.4pt height 6pt
   \kern5pt\vrule width 0.4pt}\hrule height 0.4pt}}}$}}

\begin{document}
\input psfig.tex
\newcommand{\C}{\cite}
\newcommand{\dirac}{\!\!\not{\!\partial}}
\newcommand{\Lma}{$\Lambda_{1}/M$}
\newcommand{\Lmb}{$\Lambda_{2}/M$}
\newcommand{\Eval}{$E_{\rm valence}$}
\newcommand{\Econ}{$E_{\rm cont.}$}
\newcommand{\Etot}{$E$}
\newcommand{\sval}{$\sigma_{\rm valence}$}
\newcommand{\scon}{$\sigma_{\rm cont.}$}
\newcommand{\stot}{$\sigma$}
\newcommand{\kmax}{$k_{\rm max}$}
\newcommand{\ms}{$m_{\rm s}$}
\newcommand{\rf}{$r_{f}\, M$}
\newcommand{\rg}{$r_{g}\, M$}

\normalsize\textlineskip
{\thispagestyle{empty}
\setcounter{page}{1}

\renewcommand{\thefootnote}{\fnsymbol{footnote}} 
\def\bsc{{\sc a\kern-6.4pt\sc a\kern-6.4pt\sc a}}
\def\bflatex{\bf L\kern-.30em\raise.3ex\hbox{\bsc}\kern-.14em
T\kern-.1667em\lower.7ex\hbox{E}\kern-.125em X}

\centerline{\bf STRANGENESS IN SKYRME}
\vspace*{0.035truein}
\centerline{\bf  AND IN NAMBU--JONA-LASINIO MODELS
\footnote{Talk presented at Siegen Workshop on Skyrem Model, October 1992,
to be published in the proceedings}}
\vspace{0.37truein}
\centerline{\footnotesize M. PRASZA{\L}OWICZ\footnote{
Alexander von Humboldt Fellow, on leave of absence from the
Institute of Physics, Jagellonian
University, ul. Reymonta 4, 30-059 Krak{\'o}w. Poland}}
\vspace*{0.015truein}
\centerline{\footnotesize\it Institute for Theor. Physcs II, Ruhr-University}
\baselineskip=12pt
\centerline{\footnotesize\it  4630 Bochum, Germany
}

\vspace*{0.21truein}
\abstracts{We discuss the phenomenology of SU(3) Skyrme and
Nambu--Jona-Lasinio models. We show that while the Skyrme model, in its
simplest version, is unable do reproduce the hyperon spectrum, the NJL
model describes both hyperon  and isospin  splittings with
satisfactory accuracy. The difference between the two models lies in
new {\em anomalous} moments of inertia, which vanish in the Skyrme
model and get non-zero contribution from the {\em valence} quarks in
the NJL model}{}{}

\vspace*{-3pt}\textlineskip

\textheight=8.5truein
\setcounter{footnote}{0}
\renewcommand{\thefootnote}{\alph{footnote}}

\section{Introduction}
Although QCD is now commonly accepted as the ultimate theory of strong
interactions, the low energy properties of the hadronic states are
usually calculated in terms of various {\em effective} models, which
are believed to follow from QCD.
Since there is however no rigorous direct derivation
(see {\sl e.g.}Ref.[1]),
one has some freedom in choosing ones preferable
model and preferable lagrangian.

In this talk we will discuss the phenomenology of the SU(3)
Skyrme model$^{1-5}$
and the Nambu--Jona-Lasinio (NJL) model.$^{6-9}$
In order to make the discussion transparent we will stick to the
simplest versions of the models; the more refined versions are
discussed here by other speakers.
Our attention will be focused on the splittings both in SU(3)
multiplets and in isospin multiplets. We will not consider absolute
masses which come out always too high.

We will see that in the simplest version of the Skyrme model the pattern
of splittings is generically wrong:
{}~$\Lambda-$N$=\Sigma-\Lambda=2\;(\Xi-\Sigma)$~ in contradiction with
experimental situation where ~$\Lambda-$N$\approx\Xi-\Sigma > \Sigma-\Lambda$.
To cure this disease one usually escapes to more complicated
lagrangians and, at the same time, one applies diagonalization procedure
of Yabu and Ando.$^{10}$
Although, as we will see on the example of the
explicit perturbative calculation, this method shifts the spectrum in
the right direction, one has to express a criticism against it, as
it sums up an arbitrary subseries in the strange quark masses, neglecting
other terms of the same order.

In the semibosonised  NJL model (see {\em e.g.} Ref.[11,12]),
in which quarks interact {\em via} a self-consistent meson field,
the spectrum is satisfactorily reproduced$^{6,7}$
in the linear order in $m_{\rm s}$. In the NJL model the
energy
gets contribution from the {\em valence} and {\em sea}
quarks.
The {\em classical} part, {\sl i.e.} the energy of the soliton,
is exactly the same as in the two flavor case.
The {\em quantum} corrections are
calculated by adiabatical rotation of the soliton resulting in a
hamiltonian analogous to the one of the Skyrmion.
A novelty is due to the mixed
terms linear in the current quark mass and in the rotational velocity.
These terms vanish in an ordinary Skyrme model; in the present model
they get main contribution from the {\em valence} part. The resulting
spectrum fits the data with a 10~\% accuracy. At the same time
the isospin splittings due to the $m_{\rm d}-m_{\rm u}$ mass difference
are reproduced within experimental errors.$^{9}$
Keeping in mind the
simplicity of the starting lagrangian these results are surprisingly
accurate.

\section{Gell-Mann--Okubo Mass Formulae}

Any dynamical model of light baryons has to reproduce and
also explain the Gell-Mann--Okubo$^{13,14}$
 mass formulae which are derived
assuming that the SU(3) breaking mass operator
$\Delta{\it M}$ transforms like a $Y=0,\; I=0$ and $I_{3}=0$ component
of the octet tensor operator. Then, due to the Wigner--Eckhart theorem, matrix
elements of this operator are given by:
\begin{eqnarray}
\Delta M_{B}^{(8)} =
F \left(
\begin{array}{ccl}
 8 & 8  & 8_{-} \\
000 & B  & B
\end{array}
\right)
+D \left(
\begin{array}{ccl}
 8 & 8  & 8_{+} \\
000 & B  & B
\end{array}
\right)
\label{eq:DM8}
\end{eqnarray}
for the octet, and
\begin{eqnarray}
\Delta M_{B}^{(10)} =
C \left(
\begin{array}{ccl}
 8 & 10  & 10 \\
000 & B  & B
\end{array}
\right)
\label{eq:DM10}
\end{eqnarray}
for the decuplet. The reduced matrix elements
 $F$, $D$ and $C$ are free constants.
$B=Y,\: I,\: I_{3}$ for the baryon in question.
The SU(3) Clebsch--Gordan
coefficients can be written in terms of the diagonal SU(3)
operators:$^{15}$
\begin{eqnarray}
\Delta M_{B}^{(8)}  =  -\frac{F}{2}{\rm\bf Y} -
                          \frac{D}{\sqrt{5}}
                          ( 1 - {\rm\bf I}^{2} +
                          \frac{1}{4}{\rm\bf Y}^{2}),
                    &  &
\Delta M_{B}^{(10)}  =  -\frac{C}{2 \sqrt{2}}{\rm\bf Y}. \label{eq:GO}
\end{eqnarray}

The predictive power of Eq.(\ref{eq:GO}) consists in the fact, that the
number of free parameters, which could in principle parametrize the mass
splittings, is reduced from 3 to
2 for the octet and to
1 for the decuplet.

Equations (\ref{eq:GO}) yield the relations:
\begin{eqnarray}
 & & ~~~~~~~~~~~~~~~~~~~~~~~~~~~~~~F  \, = \,  M_{\Xi} - M_{\rm N},
 \label{eq:FD} \\
 & &
\frac{1}{\sqrt{5}} D  \, = \,
\frac{1}{2} (M_{\Sigma} - M_{\Lambda})  \, = \,
\frac{1}{3} (2 M_{\Sigma} - M_{\Xi} - M_{\rm N})  \, = \,
M_{\Xi} + M_{\rm N} - 2 M_{\Lambda}, \nonumber
\end{eqnarray}
for the octet, and equal level spacing for the decuplet:
\begin{equation}
\frac{1}{2\sqrt{2}} C     \, =   \,
M_{{\Sigma}^{*}} - M_{\Delta}    \, =   \,
M_{{\Xi}^{*}} - M_{{\Sigma}^{*}} \,  =  \,
M_{\Omega} - M_{{\Xi}^{*}}.
\label{eq:C}
\end{equation}
{}From these relations we can estimate the values of parameters $F$, $D$
and $C$:
\begin{equation}
F = 379,\;\; D=79\pm17 \;\; {\rm and}\;\; C=415\pm15 \;\;
{\rm \: MeV}.
\label{eq:FDC}
\end{equation}
The resulting spectrum is presented in the first column of Fig.1.
A small admixture of other  operators like ${\bf Y}^2$
shifts the spectrum to the experimental position.

The mass splittings of the baryons belonging to the same isospin multiplet
consist of two parts: hadronic and electromagnetic,$^{16}$
namely:
\begin{equation}
\Delta m_{B} =(\Delta m_{B})_{\rm h}+(\Delta m_{B})_{\rm e}. \label{eq:DM}
\end{equation}

If the
isospin breaking is assumed to be driven by an octet isovector
tensor operator
corresponding to $I_{3}=0$ then, in analogy to the Gell-Mann--Okubo
mass formulae (\ref{eq:DM8},\ref{eq:DM10}), one gets:
\begin{eqnarray}
(\Delta m_{B}^{(8)})_{\rm h} =
 \frac{1}{\sqrt{3}} f \left(
\begin{array}{ccl}
 8 & 8  & 8_{-} \\
010 & B  & B
\end{array}
\right)
+
\sqrt{\frac{5}{3}} d \left(
\begin{array}{ccl}
 8 & 8  & 8_{+} \\
010 & B  & B
\end{array}
\right)
\label{eq:Dm8}
\end{eqnarray}
for the octet, and
\begin{eqnarray}
(\Delta m_{B}^{(10)})_{\rm h} =
\sqrt{\frac{2}{3}} c \left(
\begin{array}{ccl}
 8 & 10  & 10 \\
010 & B  & B
\end{array}
\right)
\label{eq:Dm10}
\end{eqnarray}
for the decuplet (normalization factors in front of the $f$, $d$ and
$c$  are chosen for future convenience).
Evaluating the SU(3) Clebsch-Gordan coefficients gives:$^{15}$
\begin{eqnarray}
(\Delta m_{B}^{(8)})_{\rm h}  =  -\frac{1}{3}f\: {\bf I}_{3} + d\:
{\bf Y}\: {\bf I}_{3},
                             &   &
(\Delta m_{B}^{(10)})_{\rm h}  =  -\frac{1}{3}c\: {\bf I}_{3}. \label{eq:DM
x}
\end{eqnarray}

Electromagnetic part of the isospin splittings was estimated by Gasser
and Leut{\-}wyler$^{16}$
for the octet. Their estimate confirms a reasonable
assumption that $\Sigma^{-}-\Sigma^{+}$ mass difference has no
electromagnetic contribution. This assumption allows us to determine
coefficient $f$, and also $c$ for the decuplet, where no  estimate of the
electromagnetic part of $\Sigma^{*-}-\Sigma^{*+}$
exists. Coefficient $d$ can be determined from the
hadronic part of the n--p mass difference.$^{9}$
Altogether we get:
\begin{eqnarray}
f=12.11 \pm 1.14~, &
d=1.73 \pm 0.38~~~{\rm and}  &
{}~c=6.6 \pm 1.0~~~~{\rm MeV}. \label{eq:fdc}
\end{eqnarray}

Symmetry considerations alone
are not able to provide us with any relations between
the reduced matrix elements.
Dynamical models, like Skyrme model or NJL model, make specific
predictions for these constants.
In the next sections we will calculate coefficients $F$, $D$, $C$, $f$, $d$,
and $c$ within the framework of the simplest version of the SU(3)
Skyrme model and subsequently in the simplest version of the
semibosonised SU(3) NJL model.

\section{SU(3) Skyrme Model}

Let us start by specifying the effective lagrangian
proposed by Skyrme$^{17,18}$
and later generalized by Witten:$^{19,20}$
\begin{eqnarray}
\int dt \,L_{\rm Sk} & = &\frac{F_{\pi}^{2}}{16} \int dt \, d^{3}r \,
    {\rm Tr}(\partial_{\mu}U^{\dagger}\partial^{\mu}U)
 +  \frac{1}{32e^{2}} \int dt \, d^{3}r \,
{\rm Tr}([\partial_{\mu}U\,U^{\dagger},\partial_{\nu}U\,U^{\dagger}]^{2})
\nonumber \\
& + &
\int dt\, d^{3}r\, {\rm Tr}\left\{ a\;\{U\,+\,U^{\dagger}-2\}
 + b \; \{(U\,+\,U^{\dagger})\lambda_{8}\}\right\} + N_{\rm c}\,
 \Gamma_{\rm WZ}.
\label{eq:Sklagr}
\end{eqnarray}
Here $ N_{\rm c} $  is a number of colors, parameters
$ F_{\pi} $  and $ e $, if taken from  meson  physics,
are equal to $F_{\pi} \sim 186$~MeV and  $e \sim 5.5$  respectively.
$\Gamma_{\rm WZ}$ denotes the Wess-Zumino term$^{21,22}$
and
\begin{equation}
a \, = \, \frac{F_{\pi}^{2}}{32}\,(m_{\pi}^2 + m_{\eta}^{2}), \; \; \;
b \, = \, \frac{\sqrt{3} F_{\pi}^{2}}{24}\,(m_{\pi}^2 - m_{K}^{2}).
\label{eq:ab}
\end{equation}

A {\it time-independent}
Anzatz for the SU(3) matrix $ U_{0} $ takes the form of a {\it hedgehog}:
\begin{eqnarray}
 U_{0}\, = \,\left[ \begin{array}{rcc}
    &      & 0 \\
\multicolumn{2}{c}{ \cos P(r) + i \vec{n}  \vec{\tau} \sin P(r)} &    \\
 & &     0 \\
 & &       \\
{}~~~~~ 0 & 0  & 1
\end{array} \right] ,
\label{eq:Usu3}
\end{eqnarray}
where $P(r)$ is a profile function. For the purpose of this talk we
will take
\begin{equation}
P(r) = 2\; {\rm arctan} \left( \frac{r}{r_{0}} \right)^2 \label{eq:AnzP}
\end{equation}
and calculate analytically all relevant quantities as functions of the
soliton size $r_{0}$.
 The energy of the solution (\ref{eq:Usu3}) is then given by:$^{23}$
\begin{equation}
 M_{\rm cl}  =  \frac{F_{\pi}}{e} {\pi}^{2} \frac{3 \sqrt{2}}{16}
 (4 x_{0} +\frac{15}{x_{0}}) +\frac{\mu^{2}}{e^{3} F_{\pi}} \pi^{2}
\frac{\sqrt{2}}{2} x_{0}^{3}. \label{eq:mascl}
\end{equation}
with
\begin{equation}
{\mu}^2 = \frac{ m_{\pi}^2 +2 m_{\rm K}^{2} }{3} \approx (412~~{\rm MeV})^{2},
\label{eq:mu}
\end{equation}
where $x_{0} = e F_{\pi} r_{0}$.

Minimizing Eq.(\ref{eq:mascl}) with respect to $x_{0}$ we find
in the chiral limit ($\mu^{2}=0$)~
$ x_{0}=\sqrt{15/4}$ and $M_{\rm cl}=40.54\;
F_{\pi}/e$ {\sl i.e.} approximately 1370 MeV. This  certainly
unsatisfactory result (going off chiral limit shifts  $M_{\rm cl}$
further up by
about 280 MeV) is common for all chiral models which tend to
overestimate the classical soliton mass. There is a hope that various
other effects like gluonic corrections$^{24}$
or Casimir effect$^{25,26}$
 may bring this value down to less than 1 GeV. In what follows we will
abandon the idea to fit the absolute masses, instead we will
concentrate on the mass splittings.

The splittings are calculated by rotating the static solution
\begin{equation}
 U_{0} \rightarrow A(t)\, U_{0}\, A^{\dagger}(t), \label{eq:symA}
 \end{equation}
 where $A \in SU(3)/U(1) $, since $ [\lambda_{8},U_{0}] = 0 $.
 Therefore matrix $ A $ is defined up to a {\em local} U(1) factor
 $ h= \exp(i \lambda_{8} \phi) $,{\sl i.e.} $ A $ and $Ah$ are equivalent.
 This leads to a constraint which has to be imposed on the physical
spectrum.  Introducing 8 collective  coordinates:$^{2-4}$
\begin{equation}
A^{\dagger}(t)\; \frac{d}{dt}A(t) = \frac{i}{2}  \sum_{\alpha=1}^{8}
\lambda_{\alpha} \Omega_{\alpha} ; \label{eq:adots}
\end{equation}
one gets the following quantum mechanical
lagrangian:$^{2-4}$
\begin{equation}
L =-M_{\rm cl}[P]+\frac{I_{A}[P]}{2} \sum_{i=1}^{3} \Omega_{i}^{2}
  +  \frac{I_{B}[P]}{2} \sum_{k=4}^{7} \Omega_{k}^{2}
  + \frac{N_{\rm c}}{2 \sqrt{3}} \Omega_{8}
  + \Delta m[A,P], \label{eq:Sklagra}
\end{equation}
where $\Delta m[A,P]$ corresponds to the symmetry breaking piece. The
moments of inertia can be again calculated analytically:$^{23}$
\begin{eqnarray}
I_{A} &=& \frac{1}{e^{3}F_{\pi}}\pi^{2}\frac{\sqrt{2}}{12}\,(6x_{0}^3+25x_{0})
{}~~\approx~~\frac{107}{e^{3}F_{\pi}}, \nonumber \\
I_{B} &=& \frac{1}{e^{3}F_{\pi}}\pi^{2}\frac{\sqrt{2}}{16}\,(4x_{0}^3+9x_{0})
{}~~\approx~~\frac{40.5}{e^{3}F_{\pi}}. \label{eq:inermom}
\end{eqnarray}

The standard quantization procedure$^{2-4}$
leads to the following hamiltonian:
\begin{equation}
M_{\rm B}= M_{\rm cl} + H_{\rm SU(2)} +H_{\rm SU(3)} + H_{\rm br},
\label{eq:Hamsum}
\end{equation}
with
\begin{eqnarray}
H_{\rm SU(2)}=\frac{C_{2}( {\rm SU(2)_{R}}) }{2I_{A}}, & &
H_{\rm SU(3)}=
\frac{C_{2}({\rm SU(3)_{L}})-C_{2}({\rm SU(2)_{R}})-\frac{N_{\rm c}^{2}}{12} }
{2I_{B}}, \label{eq:SkhamSU3}
\end{eqnarray}
where $C_{2}$ denotes the quadratic  Casimir operator of the
{\it right} SU(2) symmetry
corresponding to spin and of the {\it left} SU(3) corresponding to flavor.

The  wave function of the baryon state  can  be  written  as  an  SU(3)
rotation matrix:
\begin{equation}
\psi_{\rm B}(A) =  \sqrt{{\rm dim}(p,q)}\, D^{(p,q)}_{B\,S}(A)
  = \sqrt{{\rm dim}(p,q)}
\left< Y,I,I_{3} \mid D^{(p,q)}(A) \mid Y_{\rm R} ,S,-S_{3} \right>,
\label{eq:D}
\end{equation}
where quantum numbers $ B $ and $ S $ denote now hypercharge,
isospin  and  its
third component, and right hypercharge, spin and  its  third  component
(with minus sign), respectively. SU(3) representations are labeled  by
($p,q$), however not all $ p $ and $ q $ are allowed. The system is
constrained since lagrangian (\ref{eq:Sklagra})
does not contain terms quadratic in
the 8-th velocity.
The constraint $Y_{\rm R}=N_{\rm c}/3$ selects
the representations of triality zero:
\begin{equation}
8,\,10,\, \overline{10},\, 27,\, 35,\, \overline{35},\, 64,\, \ldots
\label{eq:reps}
\end{equation}
for $N_{\rm c} =3$. The success of the model is  the  prediction
that the lowest
baryonic states belong to the octet  and  decuplet  representations  of
SU(3).

The symmetry breaking Hamiltonian:
\begin{equation}
H_{\rm br}=\alpha \; D^{(8)}_{8,8}(A) \label{eq:Hbr}
\end{equation}
splits the hyperon spectrum (here representation (1,1) is denoted as (8)
and index 8 corresponds to $Y=0$ and $I=0$).
Constant $\alpha$ is given as:$^{23}$
\begin{equation}
\alpha = - \frac{\Delta\mu^{2}}{e^{3}F_{\pi}}
                  \pi^{2} \frac{\sqrt{2}}{2}\, x_{0}^{3}
{}~~\approx~~-\frac{7.7~~ {\rm GeV}^2}{e^3 F_{\pi}}, \label{eq:alpha}
\end{equation}
where we have used
\begin{equation}
\Delta\mu^{2}= \frac{2}{3} (m^{2}_{\rm K}-m^{2}_{\pi})
\approx (388~~{\rm MeV})^{2}.
\label{eq:Dmu}
\end{equation}

 \setcounter{tempfigtabc}{1}
\begin{figure}
\vspace{-6cm}
\centerline{\psfig{figure=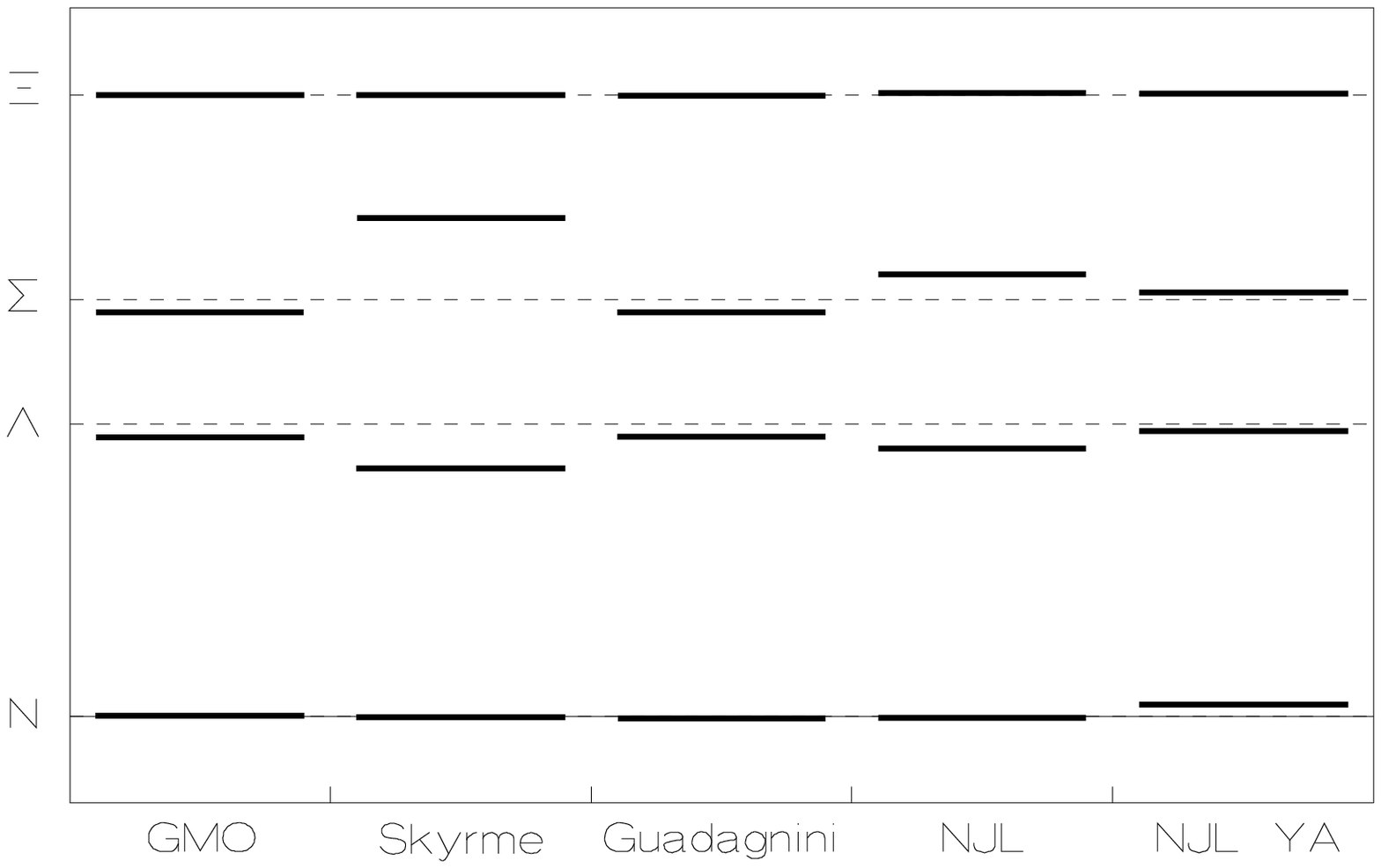,height=18 cm}}
\vspace{-7 cm}
\centerline{\psfig{figure=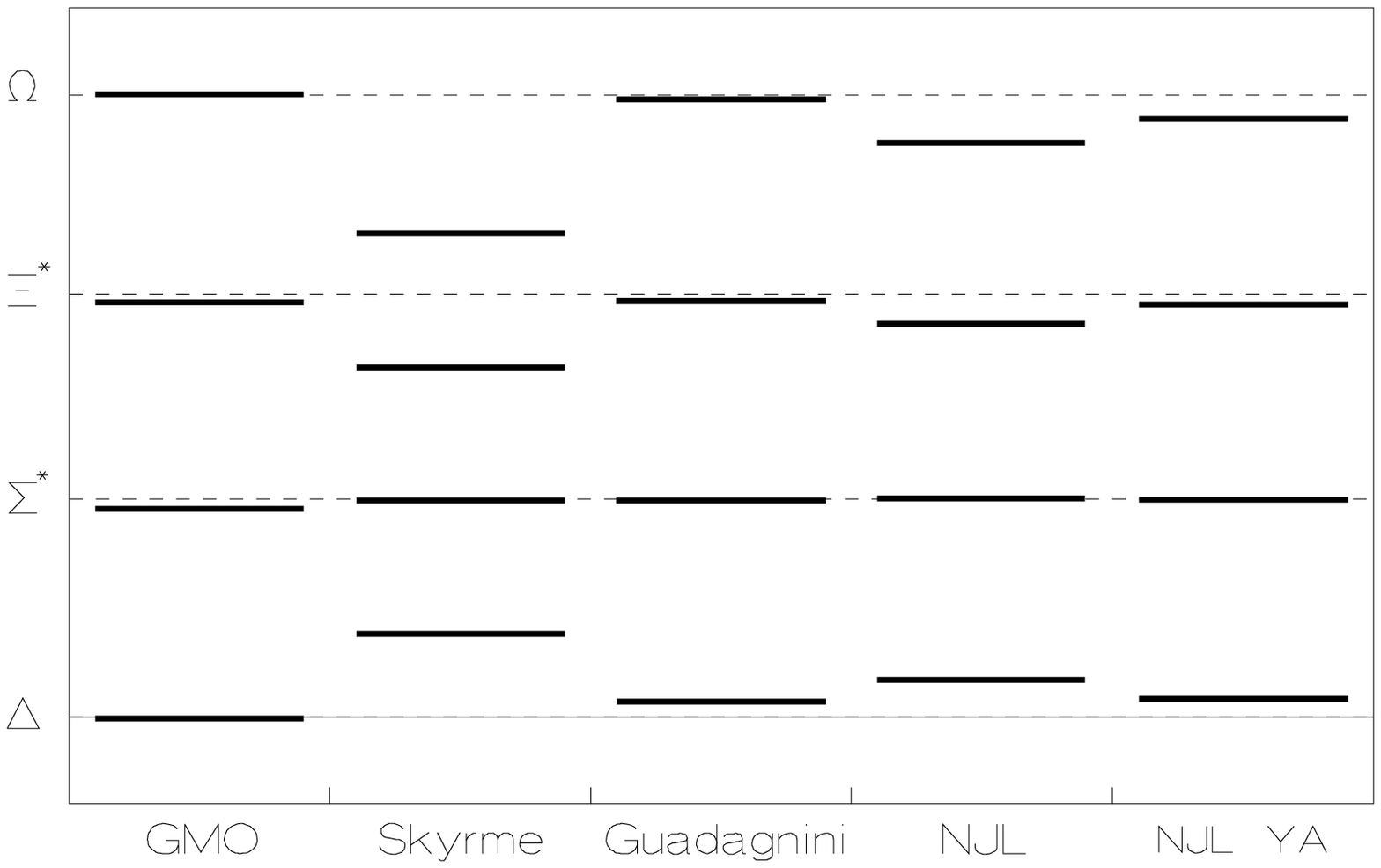,height=18cm}}
\vspace{-4 cm}
\fcaption{Spectrum of light baryons. Dashed lines represent
experimental data, solid lines correspond to theoretical predictions
described in the text.}
\end{figure}

The first and probably the only good news about the breaking
hamiltonian (\ref{eq:Hbr}) is that it automatically fulfills the
Gell-Mann--Okubo mass relation, since the {\it left} SU(3)
Clebsch-Gordan coefficients coincide with the ones of
Eqs(\ref{eq:DM8}, \ref{eq:DM10}).
The {\it right} C-G coefficients give specific
predictions for the reduced matrix elements:
\begin{equation}
F=-\frac{1}{2}\;\alpha,~~~~~D=-\frac{1}{2\sqrt{5}}\;\alpha,
{}~~~~~C=-\frac{1}{2\sqrt{2}}\;\alpha. \label{eq:FDCSk}
\end{equation}
The spectrum however looks rather odd since the two ratios:
\begin{eqnarray}
\frac{F}{D}  =  \sqrt{5}=2.24~~({\rm exp.}~~4.40),
 & &
\frac{F}{C}   =   \sqrt{2}=1.41~~({\rm exp.}~~0.91) \label{eq:ratios}
\end{eqnarray}
are far from their experimental values.
If we choose $\alpha=-758$~MeV in such a way that
$F$ is reproduced then we get
spectrum shown in the second column of Fig.1.
However $\alpha$ is not a free parameter.
The 10--8 splitting (230 MeV)
requires $e^3F_{\pi}=16.41$~GeV. Through
Eqs(\ref{eq:inermom}, \ref{eq:alpha}) we get that
$1/I_{B}=405$~MeV and
$\alpha \approx -470$~MeV, approximately 40~{\%}  smaller than the
value required to fit $F$.
One could in principle enlarge $\alpha$ by tuning the kaon mass,
but the problem of Eq.(\ref{eq:ratios}) remains.

Guadagnini$^{2}$
in the first paper on the SU(3) quantization of the Skyrme
model proposed to add {\it by hand} a term proportional to the
hyphercharge:
\begin{equation}
H_{\rm G}=\alpha \; D^{(8)}_{8,8}(A) + \beta \; {\bf Y}. \label{eq:HG}
\end{equation}
 $\beta$ is a free parameter and the reduced matrix elements read:
\begin{equation}
F=-\frac{1}{2}\;\alpha-2\;\beta,
  ~~~D=-\frac{1}{2\sqrt{5}}\;\alpha,
  ~~~C=-\frac{1}{2\sqrt{2}}(\;\alpha+8\; \beta). \label{eq:FDCG}
\end{equation}
An excellent fit to the data (column 3 in Fig.1) is obtained with
$\alpha=-382$~MeV and $\beta=-94$~MeV. This time $\alpha$ is
smaller  than the value   required by 10--8
splitting, but only by about 20~{\%}.

One way to generate {\it effective} $\beta$ is to calculate the
 $O(\alpha^2)$ correction to baryon energy
 (this approach for not too large $\alpha$  is in fact equivalent to the
diagonalization procedure of Yabu and Ando$^{10}$
which sums up the
perturbative series in $\alpha$):
\begin{eqnarray}
\Delta M_{B}^{(8)} & = &
\left(\frac{1}{4}\alpha-\frac{1}{60}\alpha^2 I_{B} \right)\;{\bf Y}
+ \left(\frac{1}{10} \alpha + \frac{2}{75} \alpha^2 I_{B} \right)
\left(1-{\bf I}^2+\frac{1}{4}{\bf Y}^2 \right) \nonumber \\
 & - &\frac{1}{750}\;\alpha^2 I_{B}\;{\bf Y}^2
 -\frac{47}{750}\;\alpha^2 I_{B}, \nonumber \\
\Delta M_{B}^{(10)} & = &
\left(\frac{1}{8}\alpha-\frac{29}{672}\alpha^2 I_{B} \right)\;{\bf Y}
-\frac{1}{168}\;\alpha^2 I_{B}\;{\bf Y}^2
 -\frac{13}{168}\;\alpha^2 I_{B}. \label{eq:E2ord}
\end{eqnarray}
Constants $F$, $D$ and $C$ can be immediately read off from
Eqs(\ref{eq:E2ord}), moreover new terms not present in the
original Gell-Mann--Okubo mass formula (\ref{eq:GO}) are generated.
Numerically
they are  small and we will neglect them in the present discussion.

In order to fit $F$ and $D$ one needs $\alpha\approx -677$~MeV and
$1/I_{B}=378$~MeV; with these values one gets a bit too small
$C=327$~MeV. These values of $\alpha$ and $I_{B}$ are again
larger than the ones required by 10--8 splitting.

This was the status of the SU(3) Skyrme model in the mid eighties.
A philosophical question whether one should trust perturbative
expansion in the strange quark mass arose.
Some people  decided to work
within an approach which breaks the SU(3) flavor symmetry already at
the level of the Anzatz for the $U_{0}$ field,$^{27,28}$
others tried to
push up $\alpha$ and $I_{B}$ either by enlarging the original Skyrme
lagrangian  or by employing
another model (like NJL model discussed in the next section).

\section{Semibosonised SU(3) Nambu--Jona-Lasinio Model}

It was already shown in the talk of Alkofer$^{12}$
that the solitonic
solutions of the semibosonised NJL model are
studied in terms of an effective action:
\begin{equation}
S_{\rm eff}=-{\rm Sp} \log (i \dirac \, -\, m\, -\, M \; U^{\gamma_{5}}).
\label{eq:Seff}
\end{equation}
This time, however, $U$ is an SU(3) matrix of Eq.(\ref{eq:Usu3}).
$M$ is
the constituent quark mass, which is in fact the only free parameter of
the model. The bare quark mass matrix
 can be written in a form:
\begin{equation}
m =
\mu_{0}\,\lambda_{0} - \mu_{8}\,\lambda_{8} -\mu_{3}\,\lambda_{3},
\label{eq:m}
\end{equation}
where $\lambda_{i}$ are Gell-Mann SU(3) matrices
($\lambda_{0}=\sqrt{2/3} \;{\bf 1}$) and
\[
 \mu_{0}  =  \frac{1}{\sqrt{6}}(m_{\rm u}+m_{\rm d}+m_{\rm s}),
{}~~\mu_{8}  = \frac{1}{\sqrt{12}}(2\,m_{\rm s}-m_{\rm u}-m_{\rm d}),
{}~~\mu_{3}  =  \frac{1}{2}(m_{\rm d}-m_{\rm u}). \]
\begin{equation}
 ~~~ \label{eq:mu083}
\end{equation}

The energy of the soliton consists of two parts: the energy of the {\em
continuum}, {\sl i.e.} the energy corresponding to the effective action
(\ref{eq:Seff}), and the energy of the {\em valence} level.
In what
follows, for simplicity, we confine our discussion to the {\em
continuum} part only, but one always has to remember that the pertinent
{\em valence} contribution has to be added.

The effective action (\ref{eq:Seff}) can be rewritten in terms of the
Euclidean spectral representation:$^{6,7}$
\begin{equation}
S_{\rm eff}=-N_{\rm c}T \, \int\frac{d\omega}{2\pi}
{\rm Tr} \log\, ( i \omega+H )
\left[ 1+\frac{1}{i\omega+H}(-i\gamma_{4}A^{\dagger} m\,A+A^{\dagger}\dot{A})
\right], \label{eq:Seff1}
\end{equation}
where H is the hermitean static hamiltonian:
$ H=\gamma_{4}(\gamma_{i}\partial_{i}+MU_{0})$, and appropriate
regularization is understood.
The static soliton solution for $H$ reduces to the one found in the
SU(2) case.  Formula
(\ref{eq:Seff1}) is already written in a form ready to be expanded in
a power series in $m$ and in generalized velocities
$\Omega$.
Let us for the moment forget about the mass matrix $m$ and expand
(\ref{eq:Seff1}) in powers of $\Omega$. We get (back in Minkowski metric)
familiar lagrangian (\ref{eq:Sklagra}) with:
\begin{eqnarray}
I_{ab}=\frac{N_{\rm c}}{4}\int\frac{d\omega}{2\pi} {\rm Tr}
\left[
\frac{1}{i\omega+H}\lambda_{a}\frac{1}{i\omega+H}\lambda_{b}
\right] = \left\{
\begin{array}{ccl}
I_{A}\delta_{ab} & {\rm for} & a,b=1...3 \\
I_{B}\delta_{ab} & {\rm for} & a,b=4...7~. \\
       0         & {\rm for} & a,b=8
\end{array}\right. & &
\label{eq:Iab}
\end{eqnarray}
The above expression is assumed to be properly
regularized and the full moments of inertia have also a {\em valence }
part.

The quantization proceeds
exactly as in the case of the Skyrme  model,$^{2-4}$
and as a result one arrives at the hamiltonian (\ref{eq:Hamsum}).

The novelty comes from the expansion in powers of the rotated matrix $m$:
\begin{equation}
L_{m} =-\sigma[\sqrt{6}\mu_{0}-\sqrt{3}(\mu_{8} D_{88}^{(8)}
           +\mu_{3} D_{38}^{(8)})]-2(\mu_{8} D_{8a}^{(8)}
           +\mu_{3} D_{3a}^{(8)})K_{ab}{\Omega}_{b},
\label{eq:Lm}
\end{equation}
where constant $\sigma$
\begin{eqnarray}
i\frac{N_{\rm c}}{4}\int\frac{d\omega}{2\pi} {\rm Tr}
\left[
\frac{1}{i\omega+H}\gamma_{4}\lambda_{a} \right] = \left\{
\begin{array}{ccl}
\sqrt{6}\;\sigma & {\rm for} & a=0 \\
 \sqrt{3}\;\sigma & {\rm for} & a=8  \\
 0  & {\rm for} & a=1...7
\end{array}\right. & &
\label{eq:sigma}
\end{eqnarray}
is related to the pion-nucleon sigma term
$\Sigma=3/2(m_{\rm u}+m_{\rm d})\sigma$ and {\em anomalous}
tensor  $K_{ab}$ is defined as:
\begin{eqnarray}
K_{ab}=i\frac{N_{\rm c}}{4}\int\frac{d\omega}{2\pi} {\rm Tr}
\left[
\frac{1}{i\omega+H}\gamma_{4} \lambda_{a}\frac{1}{i\omega+H}\lambda_{b}
\right] = \left\{
\begin{array}{cl}
K_{A}\delta_{ab} & {\rm for}~  a,b=1...3 \\
K_{B}\delta_{ab} & {\rm for}~  a,b=4...7 \\
       0         & {\rm for}~  a,b=8
\end{array}\right.. & &
\label{eq:Kab}
\end{eqnarray}
We call $K_{ab}$ {\em anomalous} since it comes from the imaginary part
of the effective action, which is related to anomaly, and as such does
not require regularization. In fact  $K_{ab}$ gets contribution almost
entirely from the {\em valence} level.

The quantized hamiltonian $H_{\rm br}$ is not as simple as in the
Skyrme model and reads:
\begin{eqnarray}
H_{\rm br} & = & \alpha_{8}\;D^{(8)}_{88}(A)+\beta_{8}\;{\bf Y}
+ \gamma_{8}\;\sum\limits_{a=1}^{3} D^{(8)}_{8a}(A) {\bf S}_{a}
\nonumber \\
&+ & \alpha_{3}\;D^{(8)}_{38}(A)+\beta_{3}\;\frac{2}{\sqrt{3}}{\bf I}_{3}
+ \gamma_{3}\;\sum\limits_{a=1}^{3} D^{(8)}_{3a}(A) {\bf S}_{a},
\label{eq:HbrNJL}
\end{eqnarray}
where
\begin{equation}
\alpha_{i}=~~\sqrt{3} \left(-\sigma +\frac{K_{B}}{I_{B}} \right) \;\mu_{i},~~~
\beta_{i}=-\sqrt{3}\frac{K_{B}}{I_{B}}\;\mu_{i},~~~
\gamma_{i}=~2 \left( \frac{K_{A}}{I_{A}} -\frac{K_{B}}{I_{B}} \right)\;\mu_{i}.
\label{eq:albega}
\end{equation}
Index $i=8$ corresponds to $Y=0,\,I=0$ and index $i= 3$ to
$Y=0,\,I=1,\,I_{3}=0$.

We see that (\ref{eq:HbrNJL}) contains a term proportional to ${\bf Y}$
which naturally arises in this model, moreover there is another term
proportional to the product of a $D$ function and the spin operator.

We adopt the following numerical procedure:
first we find the solitonic solution for a range of constituent masses
$M$, then we find the optimal value of $M$ which reproduces the
10--8 splitting due to the rotational hamiltonian $H_{\rm SU(2)}$
(\ref{eq:SkhamSU3}).
It turns out$^{6,7}$
that $M=390$~MeV and the corresponding moments of inertia
take the following values: $1/I_{A}=157$, $1/I_{B}=234$,
$1/K_{A}=469$ and
$1/K_{B}=704$~MeV, and $\sigma=3.07$.$^{29}$
We will see that not only
hyperon splittings but also isospin splittings are well reproduced. To
this end let us
let us define the following quantities:
\begin{equation}
\varphi=\sigma+2\frac{I_{B}}{K_{B}}+\frac{I_{A}}{K_{A}},~~~~~
\gamma=\sigma+2\frac{I_{B}}{K_{B}}+5\frac{I_{A}}{K_{A}},~~~~~
\delta=\sigma+2\frac{I_{B}}{K_{B}}-3\frac{I_{A}}{K_{A}}.
\label{eq:phi}
\end{equation}
Then we get:
\begin{eqnarray}
F=\frac{1}{2}\,\varphi\, m_{\rm s},~~~~~~~~~
D=\frac{1}{2\sqrt{5}}\, \delta \, m_{\rm s},~~~~~~~~~~
C=\frac{1}{2\sqrt{2}}\, \gamma\, m_{\rm s},~~~ &  &
 \label{eq:FDC-1} \\
f=\frac{3}{4}\, \varphi\, (m_{\rm d} - m_{\rm u}),~~~~
d=\frac{3}{20}\, \delta\, (m_{\rm d} - m_{\rm u}),~~~~
c=\frac{3}{8}\, \gamma\, (m_{\rm d} - m_{\rm u}).  & &
 \label{eq:fdc-1}
\end{eqnarray}

\begin{figure}
 \setcounter{tempfigtabc}{2}
\vspace{-6cm}
\centerline{\psfig{figure=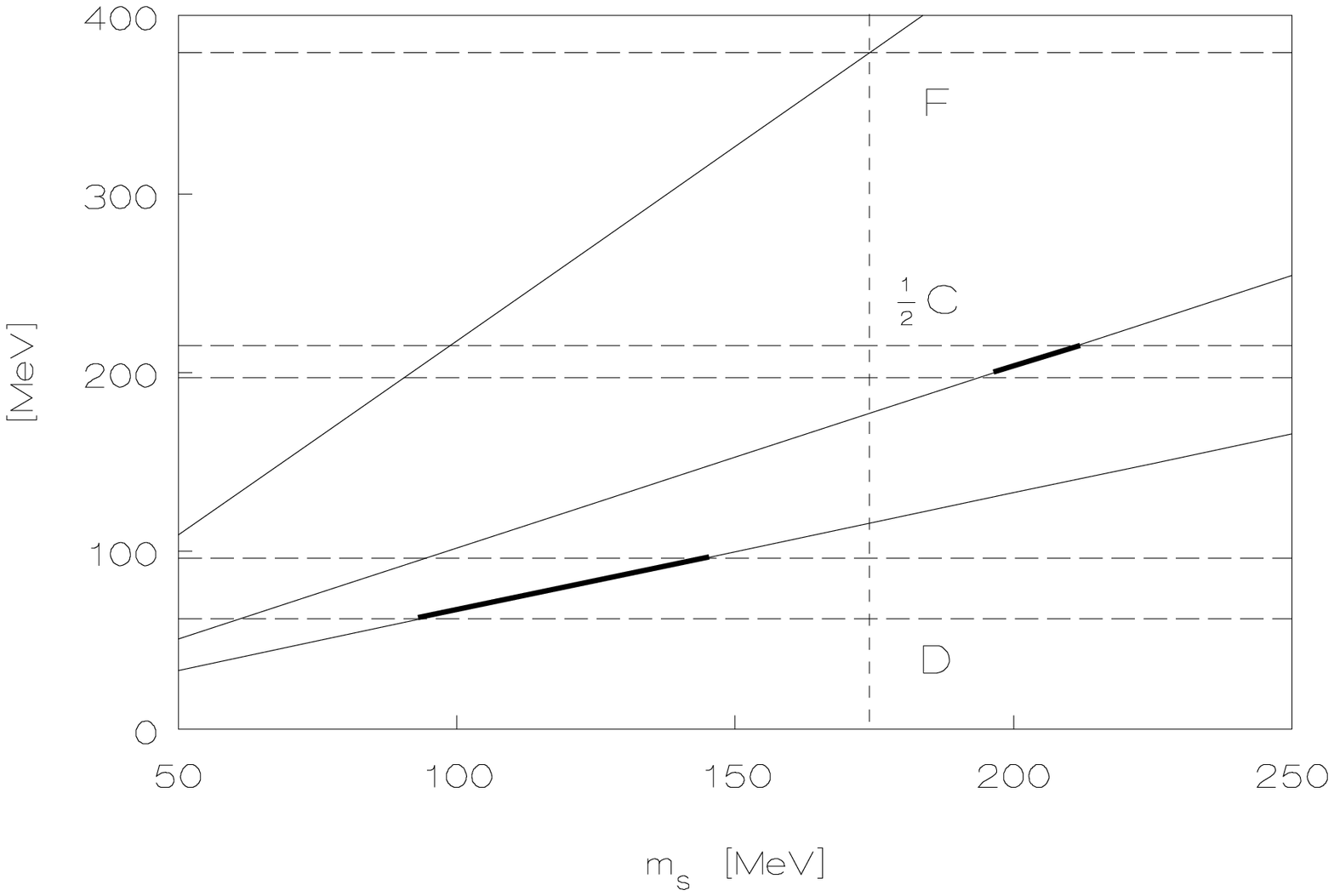,height=18cm}}
\vspace{-4cm}
\fcaption{Constants $F$, $D$ and $1/2\,C$. Solid lines represent
model predictions as functions of $m_{s    }$. Dashed
line correspond to the error bars of Eq.(6). }
\end{figure}

In Figs 2 and 3 we plot the splitting constants of
Eqs(\ref{eq:FDC-1}, \ref{eq:fdc-1}) as functions of $m_{\rm s}$ and
$m_{\rm d} - m_{\rm u}$
respectively, together with the error bars corresponding to
Eqs(\ref{eq:FDC}, \ref{eq:fdc}).
For $m_{\rm s}\approx 175$~MeV the N--$\Xi$ splitting
({\sl i.e.} constant $F$) is reproduced. For the same value of $m_{\rm s}$
constant $D$ corresponding to
$\Sigma$--$\Lambda$ splitting is overestimated by 35~MeV, whereas
constant $C$ is underestimated by 60~MeV.
The resulting spectrum is shown in column 4 of Fig.1.
On the other hand, the isospin breaking constants
$f$, $d$ and $c$ are reproduced within experimental errors for
$m_{\rm d} - m_{\rm u}\approx 3.5$~MeV (see Fig.3).

\begin{figure}
 \setcounter{tempfigtabc}{3}
 \vspace{-5cm}
\centerline{\psfig{figure=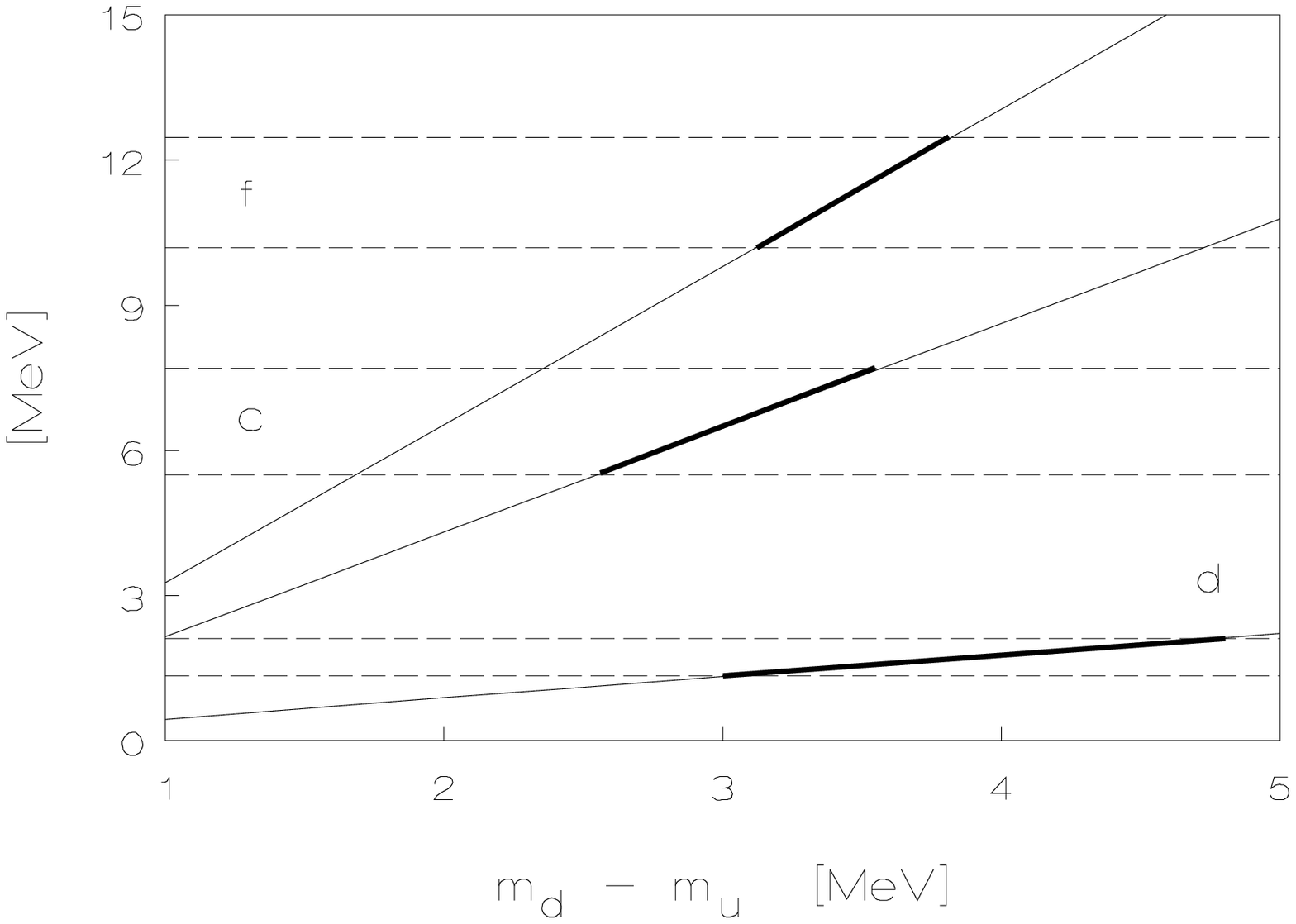,height=18cm}}
 \vspace{-4cm}
\fcaption{Constants $f$, $d$ and $c$. Solid lines represent
model predictions as functions of $m_{    d}-m_{    u}$ mass difference.
Dashed
line correspond to the error bars of Eq.(11).}
\end{figure}

The present model makes specific prediction for the ratio of hadronic
to isospin breaking constants:
\begin{equation}
\frac{f}{F}~(2.99\pm0.3)  =
\sqrt{5}\,\frac{d}{D}~(4.90\pm 2.15)   =
 \sqrt{2}\,\frac{c}{C}~(2.25\pm 0.42)~~~(\times 10^{-2}),
\end{equation}
where the numbers in brackets correspond to the experimental values of
Eqs(\ref{eq:FDC}, \ref{eq:fdc}). Certainly the central values are fairly
scattered.
We would like to offer the following explanation of this discrepancy.
The isospin splittings are proportional to a tiny parameter, namely
$m_{\rm d} - m_{\rm u}$, and therefore the first order of the
perturbation theory is legitimate. On the contrary, for the hyperon splittings
which are proportional to the much larger parameter, namely strange quark mass,
one may expect some corrections from the higher order of the
perturbative expansion in $m_{\rm s}$. Indeed,
 already the second
order brings the splittings to their experimental values with an
accuracy of a few MeV. Last column in Fig.1 shows the spectrum
obtained by applying the Yabu Ando procedure to the present model.

To summarize: we have studied the symmetry breaking effects due to the
quark masses in the Skyrme model and in the solitonic sector of the
semibosonised NJL model.
In the simplest version of the Skyrme model the splitting operator is
too simple to account  for hyperon splittings. On the contrary, in
the simplest version of the NJL model,
a satisfactory description of the hadronic mass spectrum
including both the hyperon and the isospin splittings was found.
The new terms in the splitting operator, which are not present in the
Skyrme model, are proportional to the {\em anomalous} moments of inertia
which get the main contribution from the valence level.
The absolute masses are too big, but there exist several mechanisms which
may bring them down, namely gluon corrections,$^{24}$
rotational and
translational band subtraction$^{7}$
and Casimir energies of quantum
fluctuations.$^{25,26}$

\nonumsection{Acknowledgements}

I would like to thank the organizers for very nice and stimulating
atmosphere during the Workshop.
This work summarizes a series of papers written in collaboration with
A. Blotz,
D.I. Diakonov, Z. Duli{\'n}ski, K. Goeke,
M.A. Nowak, P.O. Mazur, P.V. Pobylitsa,
and P. Sieber whom I would like to thank for many discussions.

This work was partly sponsored by {\sl Polish Research Grant 2.0091.91.01}.

\nonumsection{References}

\end{document}